\documentstyle[12pt]{article}
\def\nostrocostruttino#1\over#2{\mathrel{\mathop{\kern 0pt \rlap
{\hbox{$#1$}}} \hbox{\kern-.135em $#2$}}}
\def\sumint{\nostrocostruttino \sum \over {\displaystyle\int}}
\begin{document}
\begin{flushright}
INFNCA-TH9622 \\
DFTT 66/96 \\
hep-ph/9610388 \\
October 1996
\end{flushright}
\vspace{0.5truecm}
\renewcommand{\thefootnote}{\fnsymbol{footnote}}
\begin{center}{\large \bf Spin measurements in $lp \rightarrow hX$ deep
inelastic scattering\footnote{\,Talk delivered by J. Hansson at the
XII International Symposium on High Energy Spin Physics, Amsterdam,
Sept. 10-14, 1996.}
\\ }
\vspace {5mm}M. Anselmino$^1$, M. Boglione$^1$, J. Hansson$^{2}$ 
and F. Murgia$^3$\\
\vspace{5mm}{\small\it(1) Dipartimento
di Fisica Teorica, Universit\`a di Torino and \\
INFN, Sezione di Torino, Via P. Giuria 1, I-10125 Torino, Italy\\(2)
Department of Physics, Lule{\aa} University of Technology, S-97187
Lule\aa, Sweden\\(3) INFN, Sezione di Cagliari, Via A. Negri 18, I-09127
Cagliari, Italy }
\end{center}
\vskip 6pt
\begin{center} ABSTRACT
\vskip 5mm
\begin{minipage}{130 mm}
\small The production of hadrons in polarized lepton-nucleon deep
inelastic scattering is discussed. The helicity density matrix of the
hadron is computed within the QCD hard scattering formalism and its
elements are shown to yield information on the spin structure of the
nucleon and the spin dependence of the quark fragmentation process. The
case of $\rho$ vector mesons is considered in more detail and estimates
are given.
\end{minipage}
\end{center}

According to the QCD hard scattering scheme and the factorization
theorem \mbox{[1]-[5]} the helicity density matrix of the hadron
$h$ inclusively produced in the DIS process $\ell^\uparrow N^\uparrow
\to h^\uparrow X$ is given by %
\begin{eqnarray}
\rho_{\lambda^{\,}_h,\lambda^\prime_h}^{(s,S)}(h) &\>&
\!\!\!\!\!\!\!\!\! \frac{E_h \, d^3\sigma^{\ell,s + N,S \to h + X}}
{d^{3} {\bf p}_h} = \sum_{q; \lambda^{\,}_{\ell}, \lambda^{\,}_q,
\lambda^\prime_q} \int \frac {dx}{\pi z} \frac {1}{16 \pi x^2 s^2}
\times \\ & & \rho^{\ell,s}_{\lambda^{\,}_{\ell}, \lambda^{\,}_{\ell}}
\, \rho_{\lambda^{\,}_q, \lambda^{\prime}_q}^{q/N,S} \, f_{q/N}(x) \,
\hat M^q_{\lambda^{\,}_{\ell}, \lambda^{\,}_q; \lambda^{\,}_{\ell},
\lambda^{\,}_q} \,
\hat M^{q\textstyle{*}}_{\lambda^{\,}_{\ell}, \lambda^{\prime}_q;
\lambda^{\,}_{\ell}, \lambda^{\prime}_q} \, D_{\lambda^{\,}_h,
\lambda^{\prime}_h}^{\lambda^{\,}_q,\lambda^{\prime}_q}(z), \nonumber
\end{eqnarray}
where $\rho^{\ell,s}$ is the helicity density matrix of the initial
lepton with spin $s$, $\rho^{q/N,S}$ is the helicity density matrix of
quark $q$ inside the polarized nucleon $N$ with spin $S$ and
$f_{q/N}(x)$ is the number density of unpolarized quarks $q$ with
momentum fraction $x$ inside an unpolarized nucleon.
The $\hat M^q_{\lambda^{\,}_{\ell}, \lambda^{\,}_q; \lambda^{\,}_{\ell},
\lambda^{\,}_q}$ are the helicity amplitudes for the elementary process
$\ell q \to \ell q$. The final lepton spin is not observed and helicity
conservation of perturbative QCD and QED has already been taken into
account in the above equation. As a consequence only the diagonal
elements of $\rho^{\ell,s}$ contribute to $\rho(h)$, and non-diagonal
elements, present in case of transversely polarized leptons, do not
contribute.
$D^{\lambda^{\,}_q,\lambda^{\prime}_q}_{\lambda^{\,}_h,\lambda^{\prime}_h
}(z)$
is the product of {\it fragmentation amplitudes} %
\begin{equation}
D^{\lambda^{\,}_q,\lambda^{\prime}_q}_{\lambda^{\,}_h,\lambda^{\prime}_h}
(z) =
\sumint_{X, \lambda^{\,}_{X}}
{\cal D}_{\lambda^{\,}_{X}, \lambda_h;
\lambda^{\,}_q} \,
{\cal D}^{\textstyle *}_{\lambda^{\,}_{X}, \lambda^{\prime}_h;
\lambda^{\prime}_q}
\end{equation}
where $\sumint_{X, \lambda^{\,}_{X}}$ stands for a spin sum and
phase space integration of the \mbox{undetected} particles, considered
as a system $X$. The usual unpolarized fragmentation function
$D_{h/q}(z)$, {\it i.e.} the density number of hadrons $h$ resulting
{}from the fragmentation of an unpolarized quark $q$ and carrying a
fraction $z$ of its momentum, is given by
\begin{equation}
D_{h/q}(z) = {1\over 2} \sum_{\lambda^{\,}_q, \lambda^{\,}_h}
D^{\lambda^{\,}_q,\lambda^{\,}_q}_{\lambda^{\,}_h,\lambda^{\,}_h}(z) =
{1\over 2} \sum_{\lambda^{\,}_q, \lambda^{\,}_h}
D_{h_{\lambda^{\,}_h}/q_{\lambda^{\,}_q}}(z) \,,
\end{equation}
where $D^{\lambda^{\,}_q,\lambda^{\,}_q}_{\lambda^{\,}_h,
\lambda^{\,}_h}(z) \equiv D_{h_{\lambda^{\,}_h}/q_{\lambda^{\,}_q}}$ is
a polarized fragmentation function, {\it i.e.} the density number of
hadrons $h$ with helicity $\lambda^{\,}_h$ resulting from the
fragmentation of a quark $q$ with helicity $\lambda^{\,}_q$.

Collinear configuration (intrinsic ${\bf k}_{\perp} = 0$) together with
angular momentum conservation in the forward fragmentation process imply
\begin{equation}
D^{\lambda^{\,}_q,\lambda^{\prime}_q}_{\lambda^{\,}_h,\lambda^{\prime}_h}
= 0 \quad\quad {\mbox{\rm when}} \quad\quad \lambda^{\,}_q -
\lambda^{\prime}_q \not= \lambda^{\,}_h-\lambda^{\prime}_h \,.
\end{equation}

Eq. (1) holds at leading twist, leading order in the coupling constants
and large $Q^2$ values. The intrinsic ${\bf k}_\perp$ of the partons has
been integrated over and collinear configurations dominate both the
distribution functions and the fragmentation processes. For simplicity
of notations we have not indicated the $Q^2$ scale dependences in $f$
and $D$. The variable $z$ is related to $x$ by the usual imposition of
energy momentum conservation in the elementary 2 $\to$ 2 process. More
technical details can be found in Ref. [5].

The quark helicity density matrix $\rho^{q/N,S}$ can be decomposed as 
\begin{equation}
\rho^{q/N,S} = P_P^{q/N,S} \rho^{N,S} + P_A^{q/N,S} \rho^{N,-S},
\end{equation}
where $P_{P(A)}^{q/N,S}$ (which, in general, depends on $x$) is the
probability that the spin of the quark inside the polarized nucleon $N$
is parallel (antiparallel) to the nucleon spin $S$ and $\rho^{N,S(-S)}$
is the helicity density matrix of the nucleon with spin $S(-S)$. Notice
that
\begin{equation}
P^{q/N,S} = P_P^{q/N,S} - P_A^{q/N,S}
\end{equation}
is the component of the quark polarization vector along the parent
nucleon spin direction.

We choose $xz$ as the hadron production plane with the lepton moving
along the $z$-axis and the nucleon in the opposite direction in the
lepton-nucleon centre of mass frame. As usual we indicate by an index
$L$ the (longitudinal) nucleon spin orientation along the $z$-axis, by
an index $S$ the (sideway) orientation along the $x$-axis and by an
index $N$ the (normal) orientation along the $y$-axis.

Some elements of the helicity density matrix of the produced hadrons can
be measured via the angular distribution of the final hadron $h$ decay.
Typical examples are the $\rho \to \pi\pi$ and $\Lambda \to p \pi$
decays.

For spin-1 hadrons $(V)$ one can measure $\rho_{0,0}$ and $\rho_{1,0}$.
\\The general formulae for polarized protons and unpolarized leptons are
($T=S,N$):
\begin{eqnarray}
\rho_{0,0}^{(S_T)}(V) \, d^3\sigma &=&
\sum_q \int \frac {dx}{\pi z} \, f_{q/N} \, d\hat\sigma^q \, D_{V_0/q_+}
\\
\rho_{0,0}^{(S_L)}(V) &=& \rho_{0,0}^{(S_T)}(V) \\
\rho_{1,0}^{(S_S)}(V) \, d^3\sigma &=&
\sum_q \int \frac {dx}{\pi z} \, f_{q/N} \, {P^{q/N,S_S} \over 2} \left[
\mbox{\rm Re} \hat M_+^q \hat M_-^{q \textstyle{*}} \right] D_{1,0}^{+,-}\\
\rho_{-1,0}^{(S_S)}(V) &=& \rho_{1,0}^{(S_S)}(V)  \\
\rho_{1,0}^{(S_N)}(V) &=& - \rho_{-1,0}^{(S_N)}(V) = i
\rho_{1,0}^{(S_S)}(V) \,. \end{eqnarray}
These formulae involve the non-diagonal fragmentation functions (2).
$\hat M^q_{\pm}$ is a short notation for
$\hat M^q_{+,\pm;+,\pm}/4\sqrt {\hat s}$.
Such measurements supply information on the polarized quark
fragmentation process and the polarized distribution functions.

With $SU(6)$ wavefunctions and simple assumptions for $D_{1,0}^{+,-}$ [5] 
we get
\begin{eqnarray}
\rho_{0,0} (\rho) &=& \frac{1}{3} \\
Re \rho^{(S_N)}_{1,0} (\rho^{+,0}) & \simeq & \mbox{0.10 -- 0.15}
\end{eqnarray}
both for $\sqrt{s} =$ 23 GeV and 314 GeV, almost independently
of $p_T$ and $\mid \! x_F \! \mid$, although they have to be high enough
to justify the use of Eq. (1) and the valence quark approximation.

As we noticed after Eq. (1) only the diagonal elements of the lepton
helicity density matrix $\rho^{\ell,s}$ contribute to $\rho(h)$, so that
only longitudinal polarizations affect the results.
For longitudinally polarized leptons one obtains the same results as in
the unpolarized lepton case for the non-diagonal matrix elements and
slightly different ones for the diagonal elements. Thus, two different
measurements might yield more information.
Further discussion can be found in Ref. [5].

We also remind that according to the $SU(6)$ wavefunction the entire
$\Lambda$ polarization, which we did not discuss in detail here [5], is
due to the strange quark. Any non-zero value would offer valuable
information on the much debated issue of strange quark polarization,
$\Delta s$, inside a polarized nucleon.

\vskip 10mm
This work has been supported by the European Community under contract
CHRX-CT94-0450.

\vskip 0.2cm
\vfill{\small \begin{description}
\item{[1]} J.C. Collins, Nucl. Phys. B {\bf 394} (1993) 169.
\item{[2]} J.C. Collins, Nucl. Phys. B {\bf 396} (1993) 161.
\item{[3]} J.C. Collins, S.H. Heppelmann and G.A. Ladinsky, Nucl. Phys.
B {\bf 420} (1994) 565.
\item{[4]} For earlier work on spin asymmetries and helicity density
matrices in hard scattering see also J. Babcock, E. Monsay and D.
Sivers, Phys. Rev. D {\bf 19} (1978) 1483; \\ M. Anselmino and P. Kroll,
Phys. Rev. D {\bf 30} (1984) 36; \\ N.S. Craigie, K. Hidaka, M. Jacob
and F.M. Renard, Phys. Rep. {\bf 99} (1983) 69.
\item{[5]} M. Anselmino, M. Boglione, J. Hansson and F. Murgia, Phys.
Rev. D {\bf 54} (1996) 828.
\end{description}}
\end{document}